# Site-Percolation Threshold of Carbon Nanotube Fibers—Fast Inspection of Percolation with Markov Stochastic Theory

Fangbo Xu[1], Zhiping Xu[2] and Boris I. Yakobson[1] (✉)

[1]Department of Mechanical Engineering & Materials Science, Rice University, Houston, TX77005

[2]Department of Engineering Mechanics, Tsinghua University, Beijing 100084, China

(✉) biy@rice.edu

**Abstract**:

We present a site-percolation model based on a modified FCC lattice, as well as an efficient algorithm of inspecting percolation which takes advantage of the Markov stochastic theory, in order to study the percolation threshold of carbon nanotube (CNT) fibers. Our Markov-chain based algorithm carries out the inspection of percolation by performing repeated sparse matrix-vector multiplications, which allows parallelized computation to accelerate the inspection for a given configuration. With this approach, we determine that the site-percolation transition of CNT fibers occurs at $p_c = 0.1533 \pm 0.0013$, and analyze the dependence of the effective percolation threshold (corresponding to 0.5 percolation probability) on the length and the aspect ratio of a CNT fiber on a finite-size-scaling basis. We also discuss the aspect ratio dependence of percolation probability with various values of $p$ (not restricted to $p_c$).

# I. Introduction

Since individual carbon nanotubes (CNT) garner excellent electrical and mechanical properties,[1-6] during past decades transferring their outstanding microscopic performance to macro-scale has been an inspiring subject while various distinct routes have been developed for manufacturing fine CNT fibers.[4, 7-14]. With the current techniques, however, the produced CNT bulk is naturally a mixture of both metallic and semiconducting CNTs, with the percentage of metallic CNTs ranging roughly from 16% to 43%.[13, 15] The low purity of metallic CNTs gives rise to performance obstacles since conducting paths throughout the fiber fail to form. Hence, whether there exists and what is the critical percentage of metallic CNTs, above which the whole fiber is always conductive, become a general concern for many relevant researchers.

We say a fiber "percolates" if metallic CNTs form clusters spanning between the ends of the fiber. Percolation has been one of the most practical subjects in a wide variety of fields for over 50 years,[16-32] Our concern for the fiber is recognized as site percolation, in which each site is independently "occupied" with a probability $p$. For a regular lattice of infinite size, there exists a critical value of $p$ (percolation threshold) at which clusters of "occupied" sites that span the entire system start to appear. In practical simulations, however, with a given $p$ and the lattice structure, there are numerous configurations for arrangement of "occupied" and "unoccupied" sites, and thus one has to inspect each of them to see if it percolates to obtain the percolation probability $\Pi(p)$. Conventionally, diverse approaches derived from the Hoshen-Kopelman (HK) algorithm have been established to help check the existence of spanning clusters for a given configuration,[23, 25, 28, 33-41] but their sophisticated operation leads to fallibility in application, and consumes excessive resources by providing information irrelevant to our concern. Accordingly a more compact and efficient algorithm for inspecting connectedness, as well as an appropriate lattice model for CNT fibers, is desirable.

In this paper, we propose a modified FCC lattice to model a CNT fiber and employ the theory of finite Markov chains to direct at our purpose, so as to find out the percolation threshold of CNT fibers. Although various forms of Markov theory have been widely applied to percolation problems by setting up random fields, or characterizing percolation behavior,[42-52] it is barely used to check the occurrence of percolation. We first describe our lattice model derived from FCC, which is known as the most close-packing structure in nature. In this lattice each site is randomly chosen to be metallic with a given probability $p$, and only when the neighboring sites are both metallic is the electric current carried through. Furthermore, we implement the theory of finite Markov chains to inspect the connectedness of a particular configuration.[53, 54] This method solves the topologically complicated problem by performing simple repeated matrix-vector multiplications, which serves as one of the most important computational kernels in scientific computing and various techniques (including parallel computing) are available.[55-61] Finally, our results corroborate the finite-size scaling law, by which the percolation threshold of the lattice model for the CNT fiber, as well as its dependence on the geometry of the fiber, is obtained.

## II. Approach

### II.A The lattice model

The structure of a realistic CNT fiber is quite complicated: individual CNTs with diverse length, diameter and chirality are closely packed with numerous defects and disconnections. Apparently imperfections may raise the percolation threshold $p_c$. As the

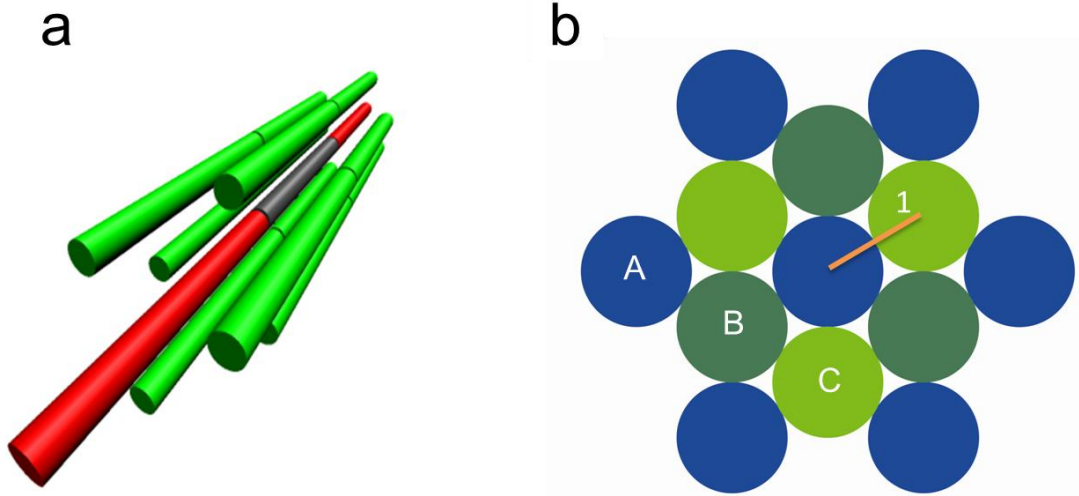

**Figure 1** (a) The schematic view of a close-packed CNT fiber. The center CNT (gray) has 2 end-contact neighbors (red) and 12 side-contact neighbors (green). (b) The cross-section of the close-packed CNT fiber. A, B and C represent the 3 distinctive layers as in a unit cell of FCC lattice. The distance between the axes of the neighboring tubes is defined as the unit of $W$, as denoted.

first step towards the reality, however, it is instructive to investigate situations free of flaws (e.g. dislocations, disconnections, vacancies) to find out the theoretical lower bound of $p_c$. Usually the cross-section of a typical CNT fiber contains tens of thousands of CNTs with similar diameters, which are closely packed in a hexagonal configuration, a way reminiscent of the (111) crystal plane of the most close-packing structures: FCC and HCP. Assuming that all the constituent CNTs share the same length and diameter, we favor the FCC structure to construct our CNT-bundle model, in which each FCC site is occupied by either a metallic or semiconducting CNT oriented along the <111> direction. As shown in Fig. 1(a), the center CNT (gray) has 14 neighbors—12 side-contact neighbors (green) and 2 end-contact neighbors (red). Figure 1(b) shows the cross-section of the fiber, while A, B and C denote the 3 layers as in a unit cell of the regular FCC lattice. Further simplification depicts CNTs as individual atoms so as to explicitly describe the geometry of the fiber. The length of the whole fiber, $L$, is defined as the number of FCC unit cells. The radius, $W$, is defined as the distance from the axis of the center CNT to that of the outmost CNTs, and the unit radius is defined as the distance between the axes of side-contact neighboring tubes, as marked in Fig. 1(b). On the other hand, if CNTs follow the HCP structure, only the A and B layers in Fig. 1(b) are present, thereby the coordinate number and the number of CNTs per unit area are both much less than FCC. Consequently HCP is not included in our percolation models.

One may note that with a given probability of being metallic, $p$, and the total number of lattice sites, $N$, the expectation value of the number of metallic sites, $N_m$, is evaluated as $\langle N_m \rangle = \sum_k k \binom{N}{k} p^k (1-p)^{N-k} = Np$, where $k$ is the number of metallic sites in each possible configuration. Therefore the probability $p$ is equivalent to the concentration of metallic sites throughout this paper. With a given $p$, the random permutation of $Np$ metallic sites is performed many times to generate distinctive configurations, and by inspecting each of them to evaluate the percolation probability $\Pi(p)$. We should state here that in theory the total sample space $\Omega$, upon which the regular probability is defined, is spanned by the total $\binom{N}{Np}$ permutations. In Monte Carlo simulations, however, we choose $Np$ sites out of $N$ with uniform probability, so that another sample space $\Omega'$ is constructed. Apparently $\Omega' \subset \Omega$, and $\Omega$ contains more configurations with inhomogeneous distribution, but $\Omega'$ is what matters in realistic experiments and serves as the sample space upon which our $\Pi(p)$ is defined throughout this paper.

**II.B Markov stochastic process**

As far as the algorithm used to inspect the percolation of a particular configuration is concerned, a naïve strategy is to carry out "depth-first-search" which searches all the possible paths constituted by the metallic sites. But the cost rapidly explodes when the metallic sites accrue, especially for a high coordinate number. In 1976, the Hoshen-Kopelman algorithm was published, providing an ingenious way for cluster analysis, and several derivatives and improvements have been developed since then. For instance, Ziff *et al.* proposed a potent algorithm which, whenever a new metallic site is added to the lattice, identifies the tree roots to which its neighbors belong by traversing respective trees and then amalgamates the trees with different roots.[23, 33, 34] This "union-find" procedure involves exploring the path leading to the root of a tree and visiting the sites along the path to make them point to the root. Deeper investigation, however, reveals this type of algorithm is not free of imperfection: (a) the "root-find" procedure has to be carried out for every visited neighbor of the current site. When it comes to another occupied site, which shares neighbors with a visited one, their common neighbors must be involved in the root-finding procedure again, and this cannot be omitted since their roots change from time to time. This becomes significant when the coordinate number is high. (b) The sites located on a "root-leading" path is repeatedly visited and modified whenever their root changes. (c) While indicating the existence of a spanning cluster, the HK-type algorithms also provide extra information such as composition of the conducting path, and the size of the clusters, which turns out irrelevant to our concern but consumes huge volume of computation and storage. Essentially, whether the spanning cluster exists or not, is equivalent to the possibility of reaching one side of the lattice from the other via the conducting bridge, which serves as nothing but the only interest of ours.

What matters is possibility, that is, zero or non-zero probability. Accordingly we employ the algorithm of finite Markov chains to calculate the probability of travel through the system. This scheme has been extensively used to check connectedness in the graph theory.[54, 62] Here we consider an electron jumping between neighboring metallic sites while its location ($X_n$) is described by the *random-walk* model characterized by the transition matrix **P**:

$$P_{ij} = P\left[X_{n+1} = j | X_n = i\right] \tag{1}$$

where $i, j$ label two metallic sites. $P_{ij}$ represents the probability of transition from $i$ to $j$. We have

$$P_{ij} = \begin{cases} 1/\deg(i), & i, j \text{ are neighbors} \\ 0, & \text{otherwise} \end{cases} \tag{2}$$

where $\deg(i)$ is the coordinate number of Site $i$. $\sum_k P_{ik} = 1$ always stands. Here transition probabilities of jumping from the current site to all the neighboring sites, are assumed even. If we suppose $i_0$ and $j_0$ mark the starting and destination sites, respectively, our percolation problem amounts to calculating the probability of hitting $j_0$ staring from $i_0$, after a certain number of steps. To this end, the well-established Markov theory has provided the answer. If we denote the set containing all the metallic sites by $S$, then we have:[62] (See Appendix)

$$\mathbf{f}^{(m)} = \begin{cases} \{P_{ij_0}, i \in S\}, & m = 1 \\ {}^{(j_0)}\mathbf{P}\mathbf{f}^{(m-1)}, & m > 1 \end{cases} \tag{3}$$

$${}^{(j_0)}P_{ik} = \begin{cases} P_{ik}, & \text{if } k \neq j_0 \\ 0, & \text{if } k = j_0 \end{cases} \tag{4}$$

where the element of $\mathbf{f}^{(m)}$ that corresponds to $i_0$, denoted by $\mathbf{f}_{i_0}^{(m)}$, represents the probability of hitting $j_0$ for the first time starting from $i_0$ after $m$ steps, and is also written as $P_{i_0}\left[\tau_{j_0}(1) = m\right]$. Note that for $m = 1$, $\mathbf{f}^{(m)}$ is simply the column of $\mathbf{P}$ which corresponds to $j_0$. If $P_{i_0}\left[\tau_{j_0}(1) = m\right] > 0$, then we are informed that after $m$ steps, the electron starting from $i_0$ is able to reach the site $j_0$, implying that there exists a conducting path connecting the two relevant sites. In this fashion, detecting percolation is solved by implementing repeated matrix-vector multiplications, as indicated by the recursive calculation of Eq. (3). Note that in most cases the transition matrix $\mathbf{P}$ is extremely sparse since all the entries are zero except those corresponding to neighboring metallic sites. On that account, a variety of techniques regarding sparse matrix-vector multiplication (even in parallel) are found available.[56, 58-61, 63-67]

In practical simulations, for a given *p*, *Np* sites are chosen to be metallic with uniform probability, and then a transition matrix **P** corresponding to this configuration is constructed in sparse format like CSR. These configurations can be assigned to respective CPU cores. When an individual core receives the job, what follows is performing repeated matrix-vector multiplications as instructed by Eq. (3) and (4), in multiple threads, until either the target probability becomes non-zero, or the maximum number of steps, $N_{max}$, is exceeded. As yet the choice of $N_{max}$ is $\min(N_b, Np)$. If $N_{max}$ is reached and the target probability still remains zero, implying that the electron cannot reach the destination after traversing all the metallic sites, or all the bonds between them, thereby the system is considered NOT to percolate. Moreover, in theory we should perform a series of matrix-vector multiplications with a distinctive $^{(j_0)}\mathbf{P}$ for each destination site $j_0$, but there might exist a great many destination sites on the end surface of the fiber if it takes a large radius. This issue can be solved by adding one more site as a "detector", say $j_1$, and assume its connection with all the metallic sites on the end surface, indicating that it takes just one more step to reach the site $j_1$ for the electron arriving at any metallic site on the end surface. By this means only one $^{(j_1)}\mathbf{P}$ needs to be constructed for a given configuration, with the price of increasing $N_{max}$ by one.

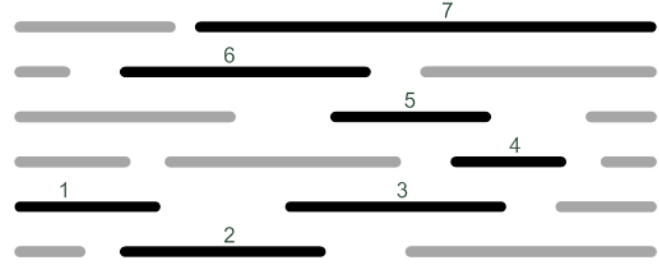

$$P = \begin{bmatrix} 0 & 1 & & & & & \\ 1/2 & 0 & 1/2 & & & & \\ & 1/2 & 0 & 1/2 & & & \\ & & 1/2 & 0 & 1/2 & & \\ & & & 1/2 & 0 & 1/2 & \\ & & & & 1/2 & 0 & 1/2 \\ & & & & & 1 & 0 \end{bmatrix}$$

$$P_1\left[\tau_7(1)=m\right] = \left[\underbrace{^{(7)}P\cdot\ldots\cdot{}^{(7)}P}_{m-1}\mathbf{f}\right](1) = \begin{cases} >0, & m=6 \\ 0, & m<6 \end{cases}$$

Figure 2 A typical example of implementation of our Markov approach for inspection of percolation for a given configuration. The top panel shows a random 2-D pattern formed by metallic tubes (dark) and insulating tubes (gray). Tube 1 and Tube 7 serve as the start and destination sites, respectively. The middle panel suggests the transition matrix $P$, where the blank sites indicate trivial elements. The column vector, f, which corresponds to the destination site, Tube 7, is marked by the dashed box. The bottom panel displays the calculation of the probability of hitting Tube 7 for the first time starting from Tube 1, after $m$ steps. $^{(7)}\mathbf{P}$ is simply identical to $\mathbf{P}$ except that the elements in the 7$^{\text{th}}$ column are all set to be zero.

Figure 2 illustrates a typical representation showing how our Markov approach works on the percolation of a random 2D pattern formed by metallic (dark) and insulating (gray) tubes. Tube 1 and Tube 7 serve as the start and destination sites, respectively. The transition matrix **P** is constructed according to Eq. (2). Surely one can assume the electron may stay at the current site for the next stage (self-loop), but that will not alter the result of percolation. The bottom panel of Fig. 2 displays the calculation of the probability of hitting Tube 7 starting from Tube 1 for the first time, after $m$ steps, which is carried out by matrix-vector multiplications, where the matrix $^{(7)}\mathbf{P}$ is readily obtained by setting the 7$^{th}$ column of $\mathbf{P}$ to be zero. If we use $\tau_{1\to 7}(1)$ to denote the number of steps it takes for the electron starting from Tube 1 to hit Tube 7 for the first time, in Fig. 2 an electron has to travel through all the bonds between metallic tubes so that $\tau_{1\to 7}(1) = N_b$. If Tube 5, for example, becomes insulating, then $\tau_{1\to 7}(1) = \infty$, and therefore the multiplication stops when $m = N_b$, implying that after walking through all the bonds it is still impossible to percolate, and therefore no more steps need to be inspected. In addition, in this way Markov method also provides the length of the shortest spanning path in the system, i.e. the length of the "backbone" of the very cluster spanning the whole lattice.

## III. Results and discussion

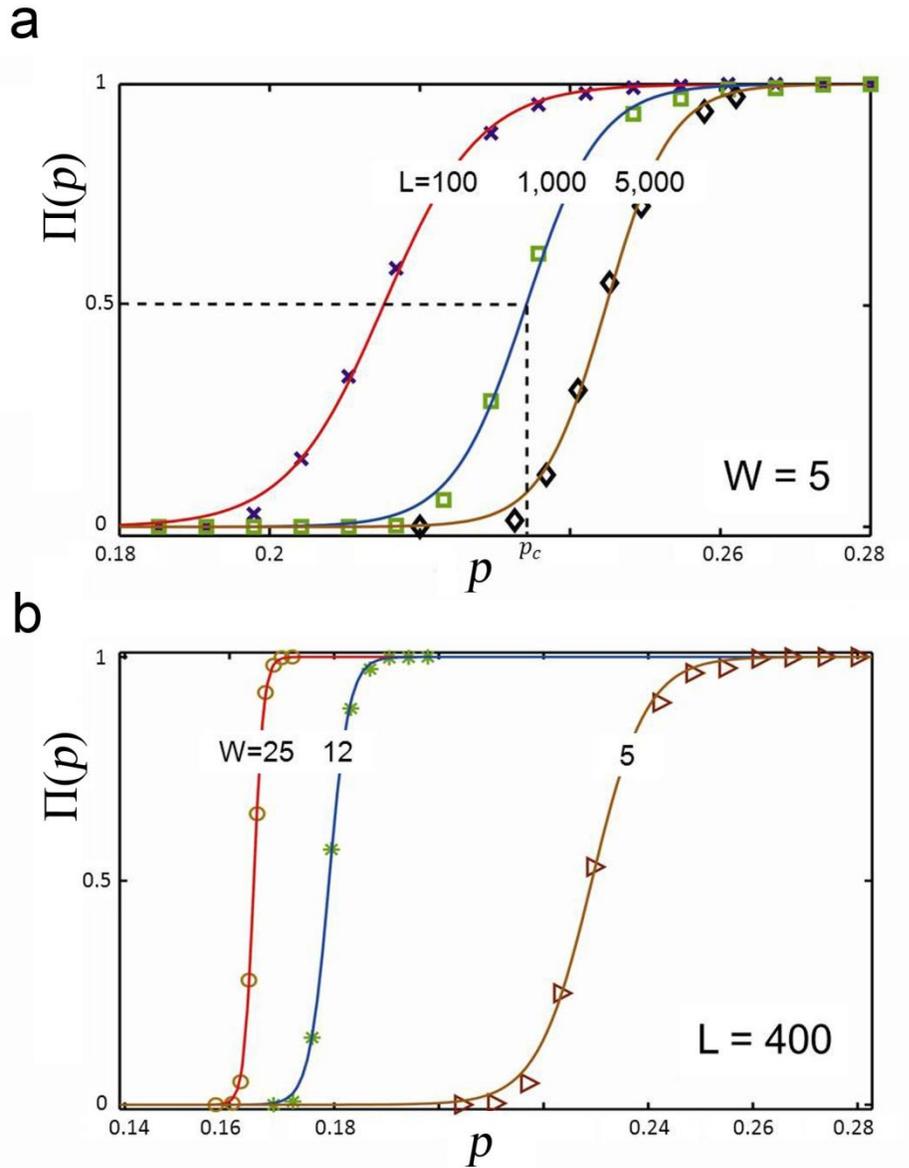

**Figure 3** The percolation probability of systems with various dimensions as a function of *p*. (a) *L*-dependence with *W* = 5. (b) *W*-dependence with *L* = 400.

Now we apply the approach described above on the FCC lattice modified for CNT fibers. Figure 3 depicts the $\Pi(p)$ in the longitudinal direction as a function of *p* for various systems with different dimensions as labeled, in which each data point is obtained by inspecting 1,024 random configurations. The largest lattice we inspected contains 2.6 million sites. Figure 3(a) illustrates the *L*-dependence of $\Pi(p)$ for a fixed $W$ while Fig. 3(b) exhibits the opposite situation. Both of the plots display the spread-out and shifting of the phase transition due to the finite-size effects, and also are found approaching step-functions when $L$ and/or $W$ advance to infinity. Faster convergence of $W$-dependence than $L$-dependence is also observed by comparison. This phenomenon can be qualitatively understood that with the same $p$, the larger $W$ the more the average number of metallic sites on a certain transection, which provides more possible paths that lead to higher probability of walking through for electrons.

In order to extrapolate the critical probability $p_c^\infty$ for infinite lattice from finite-size samples, we fit the data in Fig.3 to a variant of the Fermi-Dirac function due to the similarity of their shapes:

$$\Pi(p) = \frac{1}{1 + \exp\left(\frac{p_c - p}{T}\right)} \qquad (5)$$

where $p_c$ and $T$ are constants. One may see that as p increases, $\Pi(p)$ varies from 0 to unity, and $p_c$ is the effective percolation threshold for finite-size systems which gives $\Pi(p) = \frac{1}{2}$, as shown in Fig. 3(a). T determines the abruptness of the phase transition. According to the previous literature,[19, 23, 34, 35, 68] for a $d$-dimensional lattice of size $L^d$, using finite-size scaling law it follows that:

$$p_c = p_c^\infty + AL^{-1/\nu} \qquad (6)$$

where $A$ is a constant and the universal scaling exponent $\nu$ is 4/3 and 0.88 for 2D and 3D systems, respectively. In anisotropic systems such as hyper-rectangular cross-section bars with size $W^{d-1}L$, the aspect ratio $W/L$ also enters the expression of the scaling law. Montetti and Albano firstly studied the critical behavior of the site percolation problem on a $L \times W$ square lattice and suggest replacing the constant $A$ in Eq. (6) with a function of the aspect ratio $W/L$. [35] We follow this way and then Eq. (6) becomes:

$$p_c = p_c^\infty + C(\frac{W}{L}) \cdot L^{-1/\nu} \tag{7}$$

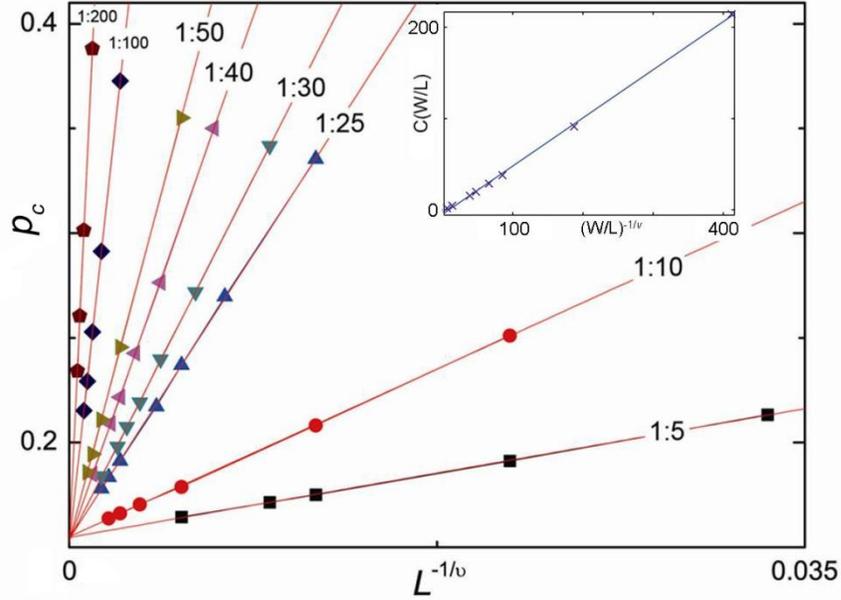

**Figure 4** The effective percolation probability of finite-size systems as a function of $L^{-1/\nu}$. The inset shows the slopes of the fitted lines in the main plot, and the straight line represents the case where $L$ and $W$ are interchangeable.

Figure 4 shows the plots of $p_c$ of our modified FCC lattice as a function of $L^{-1/\nu}$ for a variety of values of the aspect ratio *W/L*. Only those systems with the same *W/L* lie on the same straight line, and all the fitted lines share the common intercept, in agreement with Eq. (7). Furthermore, the extrapolations obtained by $L \to \infty$ give the value of the percolation threshold for the CNT fiber with infinite *L* and *W*, i.e. $p_c^\infty = 0.1533 \pm 0.0013$, lower than 0.199—the percolation threshold of the regular FCC lattice.[69] This decrease is considered reasonable due to the higher coordinate number of our lattice.

As for the slope $C(W/L)$, it can be fitted in the following form:

$$C\left(\frac{W}{L}\right) = C_0 + C_1\left(\frac{W}{L}\right)^\alpha, \quad L, W \to \infty \tag{8}$$

and then Eq. (7) becomes:

$$p_c = p_c^\infty + C_0 L^{-1/\nu} + C_1 W^\alpha L^{-\alpha-1/\nu} \tag{9}$$

Unlike the $L \times M$ square lattice, in which $L$ and $M$ are assigned identical scaling exponents due to the symmetry under the interchange of $L$ and $M$, i.e. $\alpha = -1/\nu$, our modified FCC lattice is anisotropic since we add end-end contacts only in the longitudinal direction, while electrons walk across the transection only by way of side-contacts. Therefore Eq. (9) is no longer symmetric with respect to $L$ and $W$. The inset of Fig. 4 demonstrates the slight deviation of the slopes of the straight lines in the main plot from the case in which $L$ and $W$ are interchangeable. By fitting the slopes with Eq. (8)

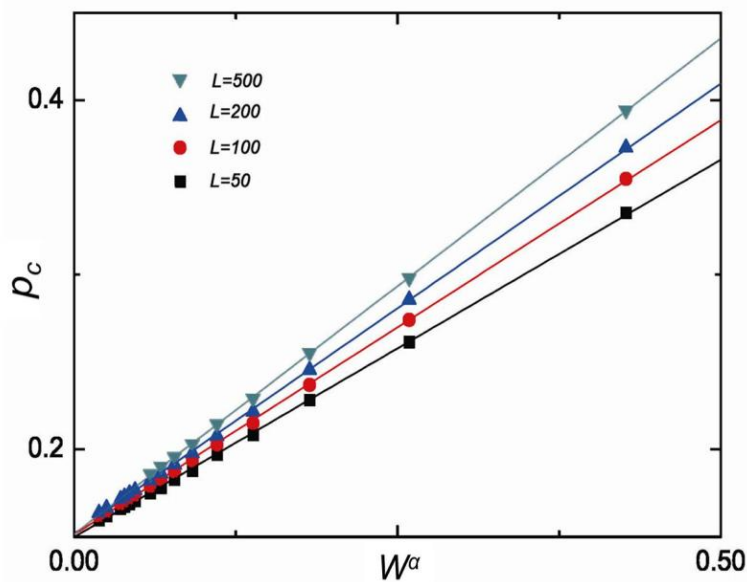

**Figure 5** The effective percolation probability versus $W^\alpha$ for various systems with different lengths.

one gets $C_0 \approx -0.83$, $C_1 \approx 0.31$, and $\alpha \approx -1.23$, with error bars of 70%, 7% and 1%, respectively, which can be considered accurate taking into account the y-data range of around 200 (cf. the inset of **Fig. 4**). Knowing these constants one can further test the validity of Eq. (9), as shown in Fig. 5, in which for a fixed length the excellent data agreement confirms the linear dependence of $p_c$ on $W^\alpha$, as well as the values of relevant coefficients with acceptable errors.

Now we obtain all of the finite-size scaling arguments of Eq. (9). Note that Eq. (9) describes the asymptotic behavior for the limit situation $L, W \to \infty$. Thus we could estimate the percolation threshold of a practical CNT fiber, say with a diameter of 1cm and a length of 1km, constituted by individual CNTs uniformly with 10nm diameter (including inter-tube distance) and 1 μm length. In this system we have $W = 10^6$ and $W/L = 1/1000$. By Eq. (9) one can find that the geometry of the system only gives a positive increment of $9.2 \times 10^{-8}$ in addition to $p_c^\infty$. Therefore one may see that at the macroscopic level, the aspect ratio does not significantly alter the percolation threshold.

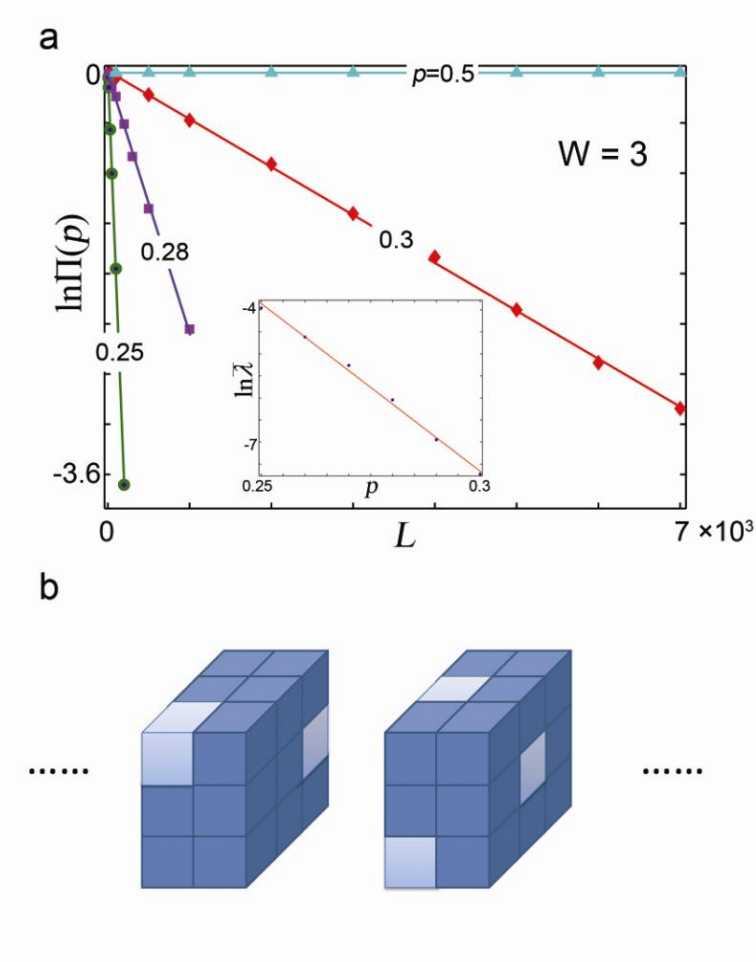

Figure 6 (a) The curves of percolation probability vs. the length of the whole lattice at various $p$ for a fixed width W=3. (b) Two immediately connected blocks of an infinitely long cube lattice with finite cross-section. The dark and light blue elements represent metallic and insulating sites, respectively. All the insulating elements are visible. The percolation probability of each block, and the connecting probability of the two adjacent blocks, can be both equal to 1, although $p$ is significantly less than 1.

Now we further the discussion about the aspect-ratio dependence of percolation probability with various values of $p$. In particular, one may speculate that for a fixed $W$, when $L \to \infty$, the fiber becomes quasi-one-dimensional so that $\Pi(p) \to 0$ unless $p = 1$. As shown in the past literature regarding a rectangular or 3D cubic system, $\Pi(p) \to 0$ as the aspect ratio advances to infinity for a given $p$ (not restricted to $p_c$).[27, 70-74] While reproducing this damping behavior, our results of simulation, however, also suggest that our lattice systems may remain percolating independent of the aspect ratio. As shown in Fig. 6(a), for a fixed $W$, when $p$ is relatively low, $\Pi(p)$ decays exponentially as $L$ increases, and the decay becomes slower for a higher $p$. If $p$ is sufficiently high, $\Pi(p)$ remains unity even when $L$ advances to infinity, but $p$ does not have to approach 100% to ensure the percolation. Moreover, in the above we have mentioned that a large $W$ may facilitate percolation, and therefore the percolation may become independent of $W/L$ at a lower $p$ if $W$ is increased. For instance, when $p = 0.3$, in Fig. 6(a) for $W = 3$ $\Pi(p)$ decays as $L$ grows; but for $W = 6$, $p = 0.3$ serves as the sufficient condition to ensure percolation for the entire range of $W/L$ (not shown).

    We understand this result in the following way. We take the whole lattice as an infinite series of linked blocks, and use $(1 - \lambda_i)$ to denote the probability of an electron jumping from the starting surface of the $i^{th}$ block to the starting surface of the $(i+1)^{th}$ block (i.e. the product of the percolation probability of the $i^{th}$ block and the connecting probability between these neighbors), and $L$ marks the number of blocks. Thus the total percolation probability of the entire lattice is evaluated as $\Pi = \prod_{i=1}^{L} (1 - \lambda_i) \sim e^{-\bar{\lambda} L}$, where $\bar{\lambda}$ is the average of $\{\lambda_i\}$.

    We could roughly estimate connection between $\bar{\lambda}$ and $p$. Referring to Eq. (5), generally if $p > p_c$ we infer $(1 - \bar{\lambda}) \sim \left(1 - e^{\frac{p_c - p}{T}}\right)$ so that $-\ln \bar{\lambda} \sim p$. The inset of Fig. 6(a) confirms this conjecture. Thus, when $p$ increases, the product of the percolation probability of each block and the connecting probability between blocks, becomes larger, equivalently $\bar{\lambda}$ diminishes exponentially. This also accounts for the fast convergence of the decaying rate as $p$ increases, as we see in Fig. 6(a).

    If each block takes small cross-section, Eq. (9) suggests that the limitation of

transversal dimensions introduces a significant increment to $p_c$. Therefore, it is likely that for quite a range of $p$, we have $\lambda_i(p) > 0$, $i \in \mathbb{N}$, leading to $\Pi \to 0$ exponentially as $L \to \infty$. On the other hand, $p = 1$ serves as too strong a condition to secure non-trivial $\Pi$. Figure 6(b) illustrates an example of cubic lattice with a side length of $W$, in which one can see that with a small $W$, even though $p$ is significantly less than 1, the $\lambda_i(p)$ of each block can be zero. Due to the finite-size effect, $\Pi(p)$ approaches unity asymptotically as $p$ increases, so that we do not expect the existence of a critical value of $p$ above which $\Pi(p)$ abruptly becomes 1 (cf. Fig. 3).

Furthermore, if we scale up each block and let $W$ approach infinity, then each block shares the same $p_c^\infty$, above which $\lambda_i(p) = 0$, $i \in \mathbb{N}$, implied by the fact that $p_c^\infty$ is also the percolation threshold of the whole lattice. Accordingly $p = p_c^\infty < 1$ serves as the sufficient condition to guarantee the occurrence of percolation independent of the aspect ratio for a macroscopic system. Note that the discussion above is made with presumption of uniform distribution of metallic and/or insulating sites among the whole lattice, that is, the sampling in each block is independently subject to the identical uniform probability distribution, and the extreme cases in which the minor insulating sites aggregate and truncate the conducting clusters are ruled out of our sample space $\Omega'$. This is the majority of situations in Monte Carlo simulations, and is also what occurs in realistic experiments.

## IV. Conclusion

In this work we propose a site-percolation model for the close-packed CNT fiber, in which individual CNTs are assumed identical and possess both end- and side-contacts leading to a coordinate number of 14. In Monte Carlo simulations, we employ the approach of finite Markov chains process to inspect percolation so as to significantly scale up the size of the lattice we study. Our numerical results agree well with the scaling functions tested with $L \times M$ square lattice,[35] and confirms that both of the length $L$ and the aspect ratio $W/L$ jointly determine the effective $p_c$ for finite-size systems. By extrapolations of these scaling functions we obtain the percolation threshold $p_c^\infty = 0.1533 \pm 0.0013$ for macroscopic systems. Due to the percolation direction and anisotropy of our lattice, the respective values of scaling exponents for $L$ and $W$ are not interchangeable. Our results also suggest that for an infinitely long fiber with a finite $W$, i.e. $W/L \to 0$, the effective $p_c$ approaches $p_c^\infty$ rather than 1. This attributes to the sample space $\Omega'$ we choose which is only spanned by configurations with uniform distribution metallic sites. As for the $\Omega$ spanned by the total permutation of metallic sites, surely $p_c \to 1$ as $L \to \infty$ since the cases in which the minor insulating sites aggregate to block the finite cross-section of the fiber are taken into account, which notwithstanding barely occur in Monte Carlo simulations as well as in realistic experiments.

**Appendix**

In this section we display the derivation of Eqs. (3)(4).[62]

Starting from Site $i$ ($X_0 = i$), after $m$ steps, the probability of hitting Site $j$ for the first time ($X_1 \neq j, X_2 \neq j, \cdots, X_{m-1} \neq j, X_m = j$) is evaluated as: (The set formed by all the metallic sites denoted by $S$)

$$\begin{aligned} f_{ij}^{(m)} &= P_i[X_1 \neq j, X_2 \neq j, \cdots, X_{m-1} \neq j, X_m = j] \\ &= \sum_{k \neq j, k \in S} P_i[X_1 = k, X_2 \neq j, \cdots, X_{m-1} \neq j, X_m = j] \\ &= \sum_{k \neq j, k \in S} P_i[X_2 \neq j, \cdots, X_{m-1} \neq j, X_m = j \mid X_1 = k] P_i[X_1 = k] \end{aligned} \quad (A.1)$$

Since in the Markov process the probability of hitting the current state only depends on the previous state, but independent of the older ones, and meanwhile the conditional probability between any two states keeps invariant in any time period, then we have:

$$\begin{aligned} &P[X_2 \neq j, \cdots, X_{m-1} \neq j, X_m = j \mid X_0 = i, X_1 = k] \\ &= P[X_2 \neq j, \cdots, X_{m-1} \neq j, X_m = j \mid X_1 = k] \\ &= P[X_1 \neq j, \cdots, X_{m-2} \neq j, X_{m-1} = j \mid X_0 = k] \\ &= f_{kj}^{(m-1)} \end{aligned} \quad (A.2)$$

From (A.2), (A.1) becomes

$$f_{ij}^{(m)} = \sum_{k \neq j} p_{ik} f_{kj}^{(m-1)}$$

Overall we have

$$f_{ij}^{(m)} = \begin{cases} p_{ij}, & \text{if } n = 1 \\ \sum_{k \neq j} p_{ik} f_{kj}^{(m-1)}, & \text{if } n > 1 \end{cases} \quad (A.3)$$

To make the computation more efficient, (A.3) can be converted into matrix-vector multiplication. That is what is given by Eqs. (3)(4).

×

Acknowledgement

# Reference

[1]	S. Sawada and N. Hamada, Solid State Commun. **83**, 917 (1992).
[2]	T. W. Ebbesen, H. J. Lezec, H. Hiura, J. W. Bennett, H. F. Ghaemi, and T. Thio, Nature **382**, 54 (1996).
[3]	M. S. Purewal, B. H. Hong, A. Ravi, B. Chandra, J. Hone, and P. Kim, Phys. Rev. Lett. **98**, 186808 (2007).
[4]	P. D. Bradford and A. E. Bogdanovich, J. Compos. Mater. **42**, 1533 (2008).
[5]	F. Xu, A. Sadrzadeh, Z. Xu, and B. I. Yakobson, J. Appl. Phys. **114**, 063714 (2013).
[6]	S. J. Tans, M. H. Devoret, H. Dai, A. Thess, R. E. Smalley, L. J. Geerligs, and C. Dekker, Nature **386**, 474 (1997).
[7]	K. Koziol, J. Vilatela, A. Moisala, M. Motta, P. Cunniff, M. Sennett, and A. Windle, Science **318**, 1892 (2007).
[8]	H. G. Chae and S. Kumar, Science **319**, 908 (2008).
[9]	N. Behabtu *et al.*, Science **339**, 182 (2013).
[10]	M. Zhang, K. R. Atkinson, and R. H. Baughman, Science **306**, 1358 (2004).
[11]	N. Behabtu, M. J. Green, and M. Pasquali, Nano Today **3**, 24 (2008).
[12]	Y. Zhao, J. Q. Wei, R. Vajtai, P. M. Ajayan, and E. V. Barrera, Scientific Reports **1**, 83, 83 (2011).
[13]	W. S. Cho, E. Hamada, Y. Kondo, and K. Takayanagi, Appl. Phys. Lett. **69**, 278 (1996).
[14]	Y. Zhao, J. Wei, R. Vajtai, P. M. Ajayan, and E. V. Barrera, Sci. Rep. **1** (2011).
[15]	S. Cambré, W. Wenseleers, E. Goovaerts, and D. E. Resasco, ACS Nano **4**, 6717 (2010).
[16]	U. Gropengiesser and D. Stauffer, Physica A **210**, 320 (1994).
[17]	D. Stauffer, Physica A **210**, 317 (1994).
[18]	D. Stauffer, J. Adler, and A. Aharony, J. Phys. A **27**, L475 (1994).
[19]	J. P. Hovi and A. Aharony, Phys. Rev. E **53**, 235 (1996).
[20]	C. D. Lorenz and R. M. Ziff, Phys. Rev. E **57**, 230 (1998).
[21]	C. D. Lorenz and R. M. Ziff, J. Phys. A **31**, 8147 (1998).
[22]	P. J. Reynolds, H. E. Stanley, and W. Klein, Phys. Rev. B **21**, 1223 (1980).
[23]	D. Stauffer, *Introduction to percolation theory* (Taylor and Francis, London, 1985).
[24]	F. Yonezawa, S. Sakamoto, and M. Hori, Phys. Rev. B **40**, 636 (1989).
[25]	J. Hoshen and R. Kopelman, Phys. Rev. B **14**, 3438 (1976).
[26]	R. M. Ziff, Phys. Rev. Lett. **69**, 2670 (1992).
[27]	D. L. Christian and M. Z. Robert, J. Phys. A: Math. Gen. **31**, 8147 (1998).
[28]	R. M. Ziff and M. E. J. Newman, Phys. Rev. E **66**, 016129 (2002).
[29]	J. C. Wierman, D. P. Naor, and R. Cheng, Phys. Rev. E **72**, 066116 (2005).
[30]	T. Kiefer, G. Villanueva, and J. Brugger, Phys. Rev. E **80**, 021104 (2009).
[31]	H. Gu and R. M. Ziff, Phys. Rev. E **85**, 051141 (2012).
[32]	S. Pfeifer, S.-H. Park, and P. R. Bandaru, J. Appl. Phys. **108** (2010).
[33]	M. E. J. Newman and R. M. Ziff, Phys. Rev. Lett. **85**, 4104 (2000).
[34]	M. E. J. Newman and R. M. Ziff, Phys. Rev. E **64**, 016706 (2001).
[35]	R. A. Monetti and E. V. Albano, Zeitschrift Fur Physik B-Condensed Matter **82**, 129 (1991).
[36]	J. Hoshen, M. W. Berry, and K. S. Minser, Phys. Rev. E **56**, 1455 (1997).
[37]	J. M. Teuler and J. C. Gimel, Comput. Phys. Commun. **130**, 118 (2000).
[38]	J. M. Constantin, M. W. Berry, and B. T. Vander Zanden, Int. J. High Perform. Comput. Appl. **11**, 34 (1997).
[39]	D. Tiggemann, International Journal of Modern Physics C **12**, 871 (2001).
[40]	M. Flanigan and P. Tamayo, Physica A **215**, 461 (1995).
[41]	N. R. Moloney and G. Pruessner, Phys. Rev. E **67**, 037701 (2003).
[42]	T. Harter, Phys. Rev. E **72**, 026120 (2005).
[43]	J. Vandenberg and C. Maes, Annals of Probability **22**, 749 (1994).
[44]	J. G. Restrepo, E. Ott, and B. R. Hunt, Phys. Rev. Lett. **100**, 058701 (2008).
[45]	N. Berger and M. Biskup, Probability Theory and Related Fields **137**, 83 (2007).
[46]	K. Bringmann, K. Mahlburg, and A. Mellit, International Mathematics Research Notices, 971 (2013).
[47]	S. Mukherjee, H. Nakanishi, and N. H. Fuchs, Phys. Rev. E **49**, 5032 (1994).

[48]   J. R. Chazottes, F. Redig, and F. Vollering, Indagationes Mathematicae-New Series **22**, 149 (2011).
[49]   Z. H. Xu and D. Han, Acta Mathematica Sinica-English Series **27**, 1813 (2011).
[50]   A. Hammond, E. Mossel, and G. Pete, Electronic Journal of Probability **17**, 1, 68 (2012).
[51]   T. Rainsford and A. Bender, IEEE Trans. Antennas Propag. **56**, 1402 (2008).
[52]   O. Haggstrom, Scand. J. Stat. **34**, 768 (2007).
[53]   S. I. Resnick, *Adventures in stochastic processes* (Birkhauser Verlag, 1992).
[54]   L. Lovász, in *Combinatorics, Paul Erdös is Eighty*, edited by J. Bolyai1993), pp. 353.
[55]   S. Pissanetzky, *Sparse matrix technology* (Academic Press, London, 1984).
[56]   Y. Saad, *Sparsekit: a basic tool kit for sparse matrix computations* (Technical Report, Computer Science Department, University of Minnesota, 1994).
[57]   K. Q. Li, Y. Pan, and S. Q. Zheng, IEEE Trans. Parallel Distrib. Syst. **9**, 705 (1998).
[58]   B. Vastenhouw and R. H. Bisseling, SIAM Rev. **47**, 67 (2005).
[59]   A. Bulu, #231, J. T. Fineman, M. Frigo, J. R. Gilbert, and C. E. Leiserson, in *Proceedings of the twenty-first annual symposium on Parallelism in algorithms and architectures* (ACM, Calgary, AB, Canada, 2009), pp. 233.
[60]   S. Williams, L. Oliker, R. Vuduc, J. Shalf, K. Yelick, and J. Demmel, Parallel Comput. **35**, 178 (2009).
[61]   G. Schubert, H. Fehske, G. Hager, and G. Wellein, Parallel Processing Letters **21**, 339 (2011).
[62]   S. I. Resnick, *Adventures in Stochastic Processes* (Birkhäuser Boston, 1992).
[63]   U. V. Catalyurek and C. Aykanat, IEEE Trans. Parallel Distrib. Syst. **10**, 673 (1999).
[64]   A. T. Ogielski and W. Aiello, Siam Journal on Scientific Computing **14**, 519 (1993).
[65]   R. S. Tuminaro, J. N. Shadid, and S. A. Hutchinson, Concurrency-Practice and Experience **10**, 229 (1998).
[66]   R. Geus and S. Rollin, Parallel Comput. **27**, 883 (2001).
[67]   E. Montagne and A. Ekambaram, Information Processing Letters **90**, 87 (2004).
[68]   E. V. Albano and H. O. Martin, Phys. Rev. B **38**, 7932 (1988).
[69]   C. D. Lorenz, R. May, and R. M. Ziff, J. Stat. Phys. **98**, 961 (2000).
[70]   J. L. Cardy, J. Phys. A: Math. Gen. **25**, L201 (1992).
[71]   R. M. Ziff, J. Phys. A: Math. Gen. **28**, 1249 (1995).
[72]   S. Tsubakihara, Phys. Rev. E **62**, 8811 (2000).
[73]   R. P. Langlands, C. Pichet, P. Pouliot, and Y. Saint-Aubin, J. Stat. Phys. **67**, 553 (1992).
[74]   R. M. Ziff, J. Phys. A: Math. Gen. **28**, 6479 (1995).